\documentclass[conference, 10pt]{IEEEtran}
\IEEEoverridecommandlockouts
\usepackage[dvipsnames]{xcolor}
\usepackage{subcaption}
\usepackage{graphicx}
\usepackage{textcomp}
\usepackage{xcolor}
\usepackage{svg}
\usepackage{float}
\usepackage{amsmath}
\usepackage{multicol}
\usepackage{array}
\usepackage{mwe}
\usepackage[noend]{algpseudocode}
\usepackage[ruled, lined, longend]{algorithm2e}
\usepackage{multirow}
\usepackage{arydshln}
\usepackage{adjustbox}
\usepackage[thinc]{esdiff}
\usepackage{booktabs} 
\usepackage{pifont}
\def\BibTeX{{\rm B\kern-.05em{\sc i\kern-.025em b}\kern-.08em
    T\kern-.1667em\lower.7ex\hbox{E}\kern-.125emX}}

\usepackage{amsfonts}
\usepackage{mathabx}

\usepackage{hyperref}

\newif\ifsubmit
\submitfalse
\ifsubmit
  \newcommand{\todo}[1]{}
  \newcommand{\ye}[1]{}
  \newcommand{\zhiheng}[1]{}
  \newcommand{\yian}[1]{}
  \newcommand{\yifan}[1]{}
  \newcommand{\yunzhe}[1]{}
  \newcommand{\sitao}[1]{}
\else
  \newcommand{\todo}[1]{{\color{red}[TODO: #1]}}
  \newcommand{\ye}[1]{{\color{red}[Ye: #1]}}
  \newcommand{\zhiheng}[1]{{\color{red}[Zhiheng: #1]}}
  \newcommand{\yifan}[1]{{\color{red}[Yifan: #1]}}
  \newcommand{\yian}[1]{{\color{red}[Yian: #1]}}
  \newcommand{\sitao}[1]{{\color{red}[Sitao: #1]}}
\fi

\begin{document}

\title{TeLLMe: An Energy-Efficient \underline{Te}rnary \underline{LLM} Accelerator for Prefill and Decode on \underline{E}dge FPGAs
}


\author{
    \IEEEauthorblockN{Ye Qiao\textsuperscript{\S}, Zhiheng Chen\textsuperscript{\S}, Yifan Zhang, Yian Wang, Sitao Huang} \\
    \textit{Department of Electrical Engineering and Computer Science} \\
    \textit{University of California,} Irvine, USA \\
    \{yeq6, zhihenc5, yifanz58, yianw11, sitaoh\}@uci.edu
    \thanks{\textsuperscript{\S}Equal contribution}
}

\maketitle

\begin{abstract}
Deploying large language models (LLMs) on edge platforms is challenged by their high computational and memory demands. Although recent low‑bit quantization methods (e.g., BitNet, DeepSeek) compress weights to as little as 1.58 bits with minimal accuracy loss, edge deployment is still constrained by limited on‑chip resources, power budgets, and the often‑neglected latency of the prefill phase. We present \textbf{TeLLMe}, the first ternary LLM accelerator for low‑power FPGAs (e.g., AMD KV260) that fully supports both prefill and autoregressive decoding using 1.58‑bit weights and 8‑bit activations. Our contributions include: (1) a table‑lookup matrix engine for ternary matmul that merges grouped activations with online precomputation to minimize resource use; (2) a fused, bandwidth‑efficient attention module featuring a reversed reordering scheme to accelerate prefill; and (3) a tightly integrated normalization and quantization–dequantization unit optimized for ultra‑low‑bit inference. Under a 7W power budget, TeLLMe delivers up to 9 tokens/s throughput over 1,024‑token contexts and prefill latencies of 0.55–1.15 s for 64–128 token prompts, marking a significant energy‑efficiency advance and establishing a new edge FPGA benchmark for generative AI.
\end{abstract}

\maketitle

\section{Introduction}

Large Language Models (LLMs) have achieved remarkable progress in recent years, powering state-of-the-art performance in natural language processing tasks such as machine translation, code generation, question answering, and conversational AI. Models like GPT-3\cite{brown2020language}, LLaMA\cite{touvron2023llama}, and DeepSeek-R1\cite{guo2025deepseek} have shown that increasing model size significantly improves generalization and task performance. However, this scaling comes with substantial costs in terms of computational demands, memory usage, and energy consumption.

Edge deployment of LLMs, i.e., running these models on low-power, resource-constrained devices such as embedded systems, FPGAs, or mobile SoCs, is a critical enabler for privacy-preserving, latency-sensitive, and autonomous applications. However, such a deployment remains challenging due to the gap between LLM complexity and the limited memory bandwidth, memory capacity, compute capacity, and power budgets on edge platforms.

To bridge this gap, recent research has focused on \textit{extreme model compression}, particularly through low-bit quantization\cite{qiao2022two,10025006}. Pioneering work such as BitNet~\cite{bitnet} demonstrated that Transformer models can be trained with 1-bit weights, while BitNet-1.58~\cite{bitnet158} and DeepSeek~\cite{deepseek} extend this to ternary quantization ($\{-1, 0, +1\}$), achieving near-parity with full-precision models. These innovations significantly reduce model size and energy cost, making LLMs more viable for edge execution.

However, deploying these compressed models on real hardware, especially FPGAs, presents unique challenges. Unlike cloud-scale GPUs, edge FPGAs have strict constraints on on-chip memory (BRAM/URAM), external DRAM bandwidth, and energy budgets. Furthermore, the requirements of autoregressive decoding (such as growing key-value (KV) caches, long context handling, and latency sensitivity) exacerbate these limitations. While most prior works focus either on model quantization or software acceleration, there is still no a systematic hardware-software co-optimization solution that fully exploit the benefits of extreme low bitwidth LLMs while meeting the computing demands of edge inference.

One critical yet often overlooked issue in edge LLM deployment 
is the disproportionate emphasis on decoding throughput, while prefill latency remains largely ignored. For example, \cite{li2025pushing} demonstrates efficient LLM decoding on embedded FPGAs but neglects the prefill stage entirely. However, prefill latency is not merely a technical detail, it is a primary bottleneck for user experience and safety in latency-sensitive edge AI applications. While prefill overhead may be negligible in cloud environments, on-device deployment places it squarely on the critical execution path. Despite its importance, prefill optimization remains significantly underexplored and demands more serious attention from the community.

To address these limitations, we present \textbf{TeLLMe}—the \textbf{Te}nary \textbf{L}arge \textbf{L}anguage \textbf{M}odel \textbf{E}dge Accelerator—the first edge FPGA-based accelerator specifically designed for ternary LLM inference with full support for both \textit{prefill} and \textit{decoding} stages. TeLLMe enables low-latency, energy-efficient deployment of LLMs on resource-constrained platforms by targeting cost-effective FPGAs such as AMD KV260. It supports ternary-quantized weights (1.58-bit) and 8-bit activations.

Our design co-optimizes compute, memory, and scheduling efficiency, key contributions are as follows:
\begin{itemize}
    \item We develop the first end-to-end edge FPGA accelerator for ternary LLMs supporting both prefill and decoding stages.
    \item We propose Table-lookup-based ternary matmul, an efficient and resource-saving matrix multiplication unit that specially optimized for FPGA, reusing grouped activations and online computation for ternary matmuls across projection and FFN layers.
    \item We introduce a fused attention unit for prefill and decoding, incorporating a novel reversed attention mechanism and fully fused pipeline to minimize off-chip data movement, avoid redundant masked computation, and guarantee the parallelism at the same time.

    \item We deliver up to 9.51 tokens/sec generation in up to 1024 token contexts while consuming less than 7W power, outperforming mobile SoCs with much lower power budgets.
\end{itemize}

TeLLMe achieves a prefill latency from 0.55s to 1.15s for prompt sizes of 64 to 128 tokens and delivers up to \textbf{9.51 tokens/s} decoding throughput with support for \textbf{1024-token context lengths} on edge FPGAs, all while operating under \textbf{7 watts} of power consumption. This marks a significant advancement over existing mobile-edge devices and prior FPGA-based accelerators, which typically require higher precision and greater power budgets.

To the best of our knowledge, TeLLMe is the first accelerator to provide end-to-end support for ternary LLM inference—including both prefill and decoding stages—on real FPGA hardware, establishing a new baseline for energy-efficient, low-latency generative AI at the edge.

\section{Background and Related Work}
\subsection{LLM Basic}

A typical LLM, such as LLaMA, is composed of multiple identical transformer blocks, each containing an attention module followed by a multilayer perceptron (MLP) module, as illustrated in Fig. \ref{Fig:Bitnet}. Within each attention module, three linear projections are used to compute the query (Q), key (K), and value (V) representations. These are then processed by a multi-head attention mechanism that incorporates both the current QKV tensors and historical KV cache. The MLP module consists of an up-projection and down-projection layer, along with an additional gating projection applied to the up-projection output.

The generative inference of LLMs is typically divided into two stages: the \textbf{prefill phase} and the \textbf{decode phase} (generation), as shown in Figure \ref{Fig:Bitnet}. During the prefill phase, the entire prompt is processed through the full model stack to produce the first output token. This phase involves parallel computation across multiple input tokens and is dominated by compute-intensive matrix–matrix multiplications, particularly within the linear projection layers. In contrast, the decode phase proceeds in an autoregressive fashion, generating one token at a time by feeding the previously generated token back into the model. This phase is typically memory-bound due to its reliance on KV cache lookup and smaller matrix–vector operations.

Following the observations in Chen et al.\cite{chen2024understanding} , while FPGAs are generally less efficient than GPUs during the compute-heavy prefill stage, they exhibit competitive advantages during the memory-intensive decode phase. In this work, we prioritize optimizing the decode phase of LLM inference to fully leverage the strengths of FPGA architectures.

\subsection{Binary, Ternary, and Low-Bit Quantized Transformers}

Model quantization is a key technique for compressing LLMs to run on constrained devices. Most conventional quantization approaches target 8-bit or 4-bit representations, but recent work has pushed the boundary down to the binary regime.

\textbf{BitNet}~\cite{bitnet} introduced a method for training Transformers from scratch using 1-bit weights. Despite extreme quantization, BitNet maintained competitive perplexity through custom scaling and layer-wise normalization strategies. Building on this, \textbf{BitNet-1.58}~\cite{bitnet158} introduced ternary weight representations ($\{-1, 0, +1\}$), striking a balance between expressiveness and compression. Both approaches highlight the potential of binary LLMs in terms of storage, throughput, and energy efficiency. Similarly, \textbf{FBI-LLM}~\cite{fbillm} and \textbf{OneBit}~\cite{onebit} demonstrate fully binarized models trained using autoregressive distillation, achieving promising results on open-domain generation tasks. \textbf{DeepSeek-R1}~\cite{deepseek} presents a hybrid quantization strategy applying ternary quantization to Mixture-of-Expert (MoE) layers, achieving up to $80\%$ model size reduction on a 671B model. Beyond training-time quantization, post-training quantization (PTQ) also plays a role. \textbf{BitDistiller}~\cite{bitdistiller} combines self-distillation with quantization-aware techniques to push 3-bit and 2-bit LLM performance to new levels. \textbf{QuIP}~\cite{quip} introduces 2-bit quantization with incoherence processing and rounding guarantees.

These works focus on algorithmic aspects. In contrast, \textbf{TeLLMe} provides a hardware-aligned solution for deploying such models in practice, offering both matmul reuse and pipeline fusion for edge execution.

\subsection{Edge-Focused LLM Acceleration on FPGA}

Deploying Transformers on FPGAs is challenging due to limited bandwidth and logic resources. Several works have explored quantized Transformer accelerators on embedded FPGAs.

\textbf{T-MAC}~\cite{tmac} implements a table-lookup-based (TL-based) matrix multiplication (matmul) kernel for CPUs using low-bit weights and high-bit activations. It achieves notable performance on Apple M2 and Raspberry Pi 5, but being software-based, it lacks the deep hardware-level optimization required for maximum efficiency. 

\textbf{{Li et al.}}~\cite{li2025pushing} successfully implemented a 4-bit quantized LLaMA2-7B model on the AMD KV260 platform. Although the model weights are quantized to 4-bit, the decoding computations rely on unquantized FP16 activations, thereby requiring all operations to be conducted in FP16 and preventing the use of more efficient 4-bit computation units. Moreover, hardware acceleration is limited to the decoding stage and does not address the computational demands of the prefilling stage. 

\textbf{LlamaF}~\cite{llamaf} targets LLaMA2-style models with int8 quantization on ZCU102. It leverages pipelined matrix-vector units and asynchronous scheduling but does not address the demands of long-context decoding and ignore prefill state entirely.

\textbf{Edge-MoE}~\cite{edgemoe} introduces a memory-efficient MoE vision transformer accelerator using dynamic task-level sparsity. 
A key technique is the specialized reordering to enable the data reuse of attention computation, which is the idea we extend in TeLLMe’s prefill module.

\textbf{SECDA}~\cite{haris2024designing} SECDA designs the MatMul accelerator supporting block floating-point quantized operations on PYNQ, reducing latency by 11x compared to dual-core Arm NEON-based CPU execution for the TinyLlama model. However, the token per second is only 0.58, which means 2 seconds for one token generation.


Compared to these, our work is the first to unify binary weight inference, prefill/decoding support, and FPGA-level memory hierarchy optimization into one cohesive design.


\begin{figure}[t] 
\centering
\includegraphics[width=\linewidth]{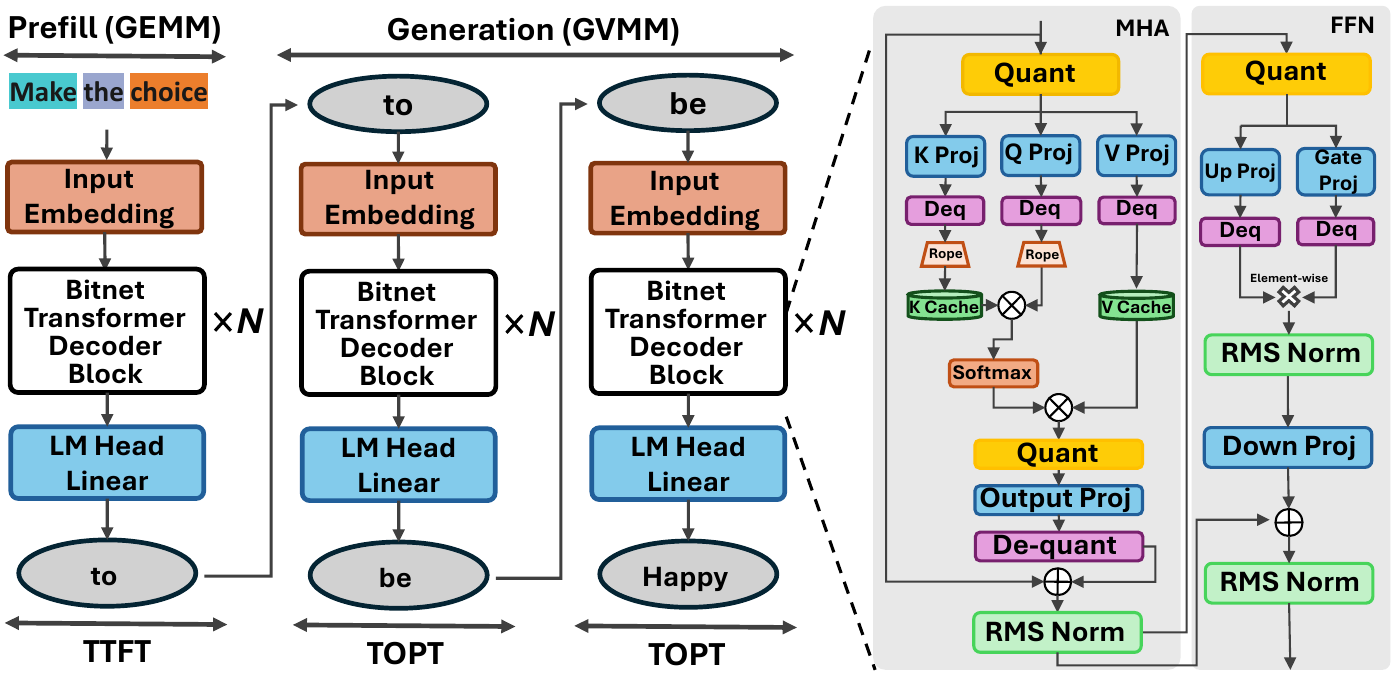}
\caption{Breakdown of TeLLMe 1.58-bit Model Inference Process with Prefill and Generation}
\label{Fig:Bitnet}
\end{figure}

\section{TeLLMe Hardware Design}
\begin{figure}[t] 
\centering
\includegraphics[width=\linewidth]{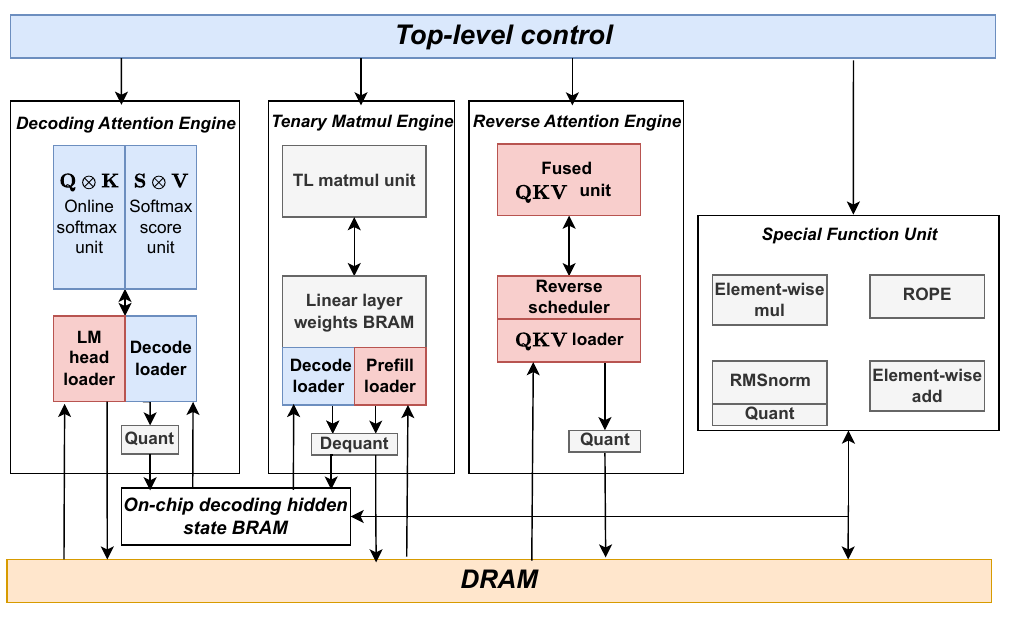}
\caption{System architecture of TeLLMe.}
\label{fig:sys_arch} 
\end{figure}

As shown in Fig. \ref{fig:sys_arch}, the accelerator design primarily consists of the following modules: (1) a table-look-up-based ternary matrix multiplication for both the decoding and prefill passes; (2) a specialized reverse reorder for prefill attention; (3) a unified decoding attention and language model head (LM Head); and (4) specialized functional units.

\subsection{Table-lookup-based Ternary MatMul on FPGA}
Table-lookup-based (TL-based) matrix multiplication (matmul) is a highly efficient method for ternary matmul in CPUs, leveraging specialized ARM NEON and AVX instructions for 128/256-bit table look-up operations. However, its limited table size often incurs frequent memory accesses and increased latency \cite{tmac}. FPGAs offer an ideal solution by exploiting their intrinsic lookup table units (LUT) resources to support larger tables, yet prior FPGA works \cite{TLMAC} primarily focus on module and design automation level optimizations rather than comprehensive dataflow or scheduling strategies. In this work, we present a TL-based ternary matmul implementation on FPGAs, coupled with an in-depth exploration of efficient pipeline scheduling to maximize performance.

\subsubsection{Algorithm background}


In general, the bit-wise operation for mixed-precision matrix multiplication can be expressed as follows:
\begin{align}
    {\bf A} \otimes {\bf W} = {\bf A} \otimes \left(\sum_{i = 0}^{n-1} 2^i {\bf W}_{i}\right) = \left(\sum_{i = 0}^{n-1} 2^i {\bf A} \otimes {\bf W}_{i}\right)
\end{align}

Considering the ternary matrix multiplication with \( W \in \{-1, 0, 1\} \), the above equation can be simplified to:

\begin{align}
    {\bf A} \otimes {\bf W} = {\bf A} \otimes {\bf W}_{0}, \quad {{\bf W}_{\text{ternary}}} \in \{-1, 0, 1\}
\end{align}
In this case, simple summation and subtraction can replace the multiplication process. However, the method of selecting -1, 0, and 1 to determine the summation and subtraction may not be optimal, as there are only a limited number of combinations of -1, 0, and 1, leading to repetitive computations for the corresponding \( \bf A \) entries. Furthermore, when increasing computation parallelism by duplicating the selection unit of the adding and subtracting path, the resource consumption of the selection may exceed that of the TL tables themselves. This is because the multiple reading ports of the on-chip distributed RAM unit can support multiple accesses to the TL tables, requiring only additional buffers for addressing. The supportive ablation study will be presented in the next subsection.

\begin{algorithm}
\tiny

\caption{TL-based Ternary Matmul}
\label{alg:tmac_stream}
\SetKwInOut{Input}{Input}
\SetKwInOut{Output}{Output}

\Input{
    $\bf \bf A$: Input activation stream (shape $[M][N]$)\;
    ${\bf W}_{idx}$ = Offline\_preprocess(${\bf W}$): Offline-preprocessed weight indices  (shape $[N/(T*G)][K]$)\;
}
\Output{
    $\bf O$: Output activation stream (shape $[M][K]$)\;
}
\BlankLine
Initialize:\;
\Indp
    $\mathit{TL\_TABLE}[N][3^{G}] \gets 0$ \tcp*[r]{Table for all signed combinations}
    $\mathit{A\_BLOCK}[T \times G ] \gets 0$ \tcp*[r]{Activation buffer}
    $\mathit{O\_BLOCK}[K] \gets 0$ \tcp*[r]{Output vector accumulator}
\Indm

\For{$i \gets 0$ \KwTo $M - 1$}{
    \For{$j \gets 0$ \KwTo $N - 1$ \textbf{step} $T \times G$}{
        \tcp{Load activation block}
        \For{$p \gets 0$ \KwTo $T \times G - 1$}{
            $\mathit{ A\_BLOCK}[p] \gets \text{A.read()}$\;
        }
        
        \tcp{Set up values of TL\_TABLE}
        \For{$t \gets 0$ \KwTo $T - 1$}{
            $\mathit{val}_{1\dots G} \gets \mathit{A\_BLOCK}[t \times G : (t + 1)\times G  -1]$\;
         \text{TL\_TABLE\_set\_up}$(\mathit{val}_{1\dots G})$\;
        }
        
        \tcp{Process hidden dimension}
        \For{$m \gets 0$ \KwTo K \textbf{step} $Q$}{
            \For{$n \gets 0$ \KwTo $Q - 1$}{
                $\mathit{idx\_vec} \gets B\left[\left\lfloor\frac{j}{T \times G}\right\rfloor\right][m + n]$\;
                \For{$t \gets 0$ \KwTo $N - 1$}{
                    $\mathit{TL\_TABLE\_idx} \gets \mathit{idx\_vec}[t]$\;
                    $\mathit{O\_BLOCK}[m + n] \gets \mathit{O\_BLOCK}[m + n] + \mathit{TL\_TABLE}[t][\mathit{TL\_TABLE\_idx}]$\;
                }
            }
        }
    }
    
    \tcp{Write output}
    \For{$p \gets 0$ \KwTo $K - 1$}{
        $\text{{\bf O}.write}(\mathit{C\_BLOCK}[p])$\;
        $\mathit{O\_BLOCK}[p] \gets 0$\;
    }
}

\BlankLine
\SetKwProg{Fn}{Function}{:}{}
\Fn{\text{Offline\_preprocess}($\bf W$)}{
    \Return Encode every $G$ value as an index in the matrix and pack every $T$ values as a index\_vector $idx\_vec$\;
}
\BlankLine
\SetKwProg{Fn}{Function}{:}{}
\Fn{\text{TL\_TABLE\_set\_up}($\mathit{val}_{1\dots G}$)}{
    \Return return all $3^G$ add and subtract combination\;
}
\end{algorithm}

\begin{figure}[t] 
\centering
\includegraphics[width=\linewidth]{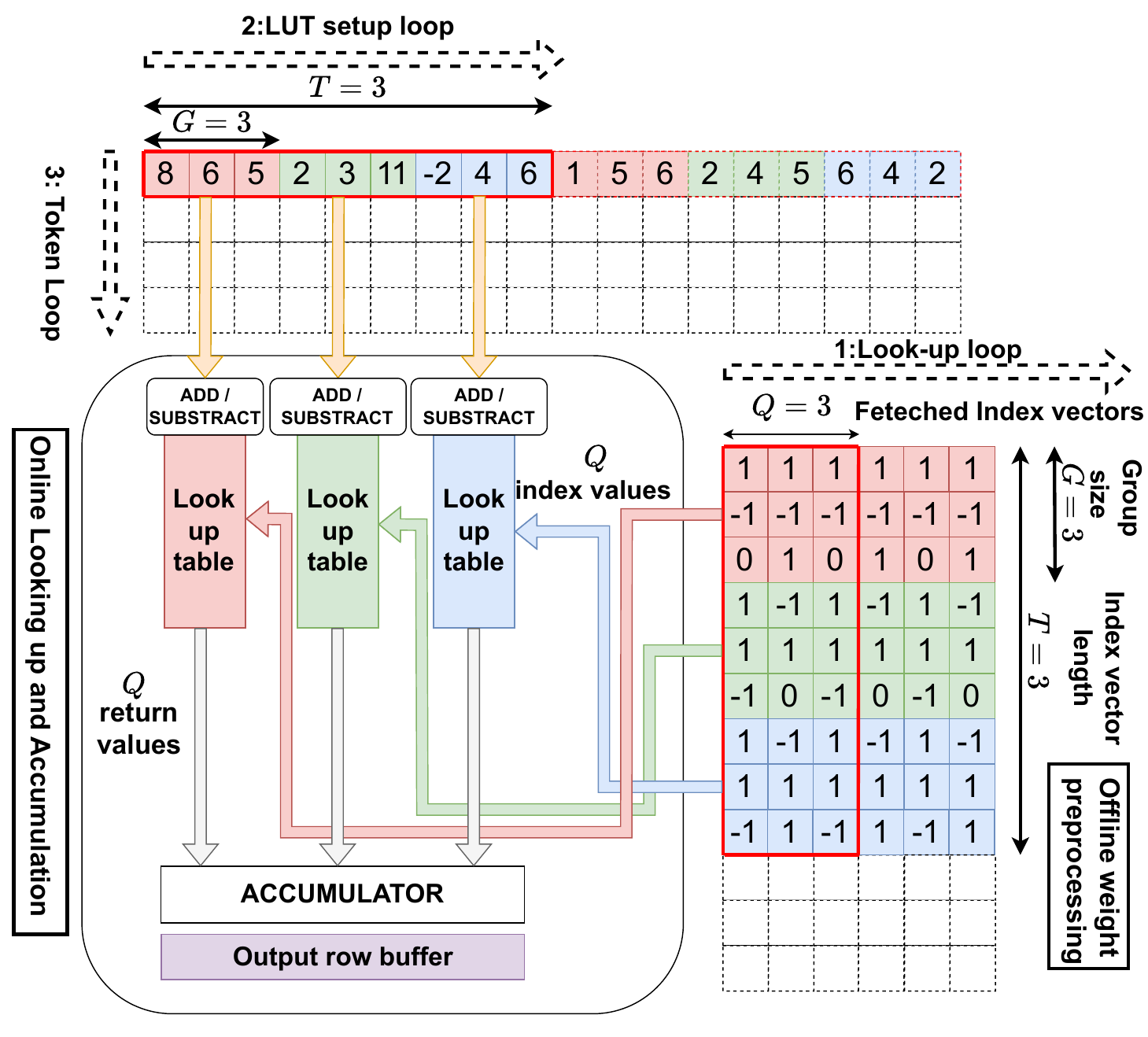}
\caption{Dataflow and architecture of TL-based ternary matMul ($G = 4$)}
\label{fig:lutmatmul} 
\end{figure}


As described in \textbf{Algorithm} \ref{alg:tmac_stream} and Fig. \ref{fig:lutmatmul}, the TL-based matmul can be divided into two stages: (1) preprocessing the weights into groups sized $G$, and (2) performing the online ternary matrix multiplication computation. 

In the preprocessing stage, assume that every $G = 3$ ternary values are packed into a single index for TL table addressing, resulting in \(3^{G} = 3 \times 3 \times 3 = 27\) combinations. The index representation for this packing requires \( \log_2 27 \approx 5 \) bits. Let \( {\bf A} \in \mathbb{N}^{M \times N} \) and \({\bf W} \in \text{ternary}^{N \times K} \). The preprocessing of the weights involves encoding every group of \( G = 3 \) ternary values into a 5-bit packed index.

In the online stage, we perform vector-wise tiled matrix multiplication, where \( {\bf A} \) and \( \bf B \). The first step is to establish the TL table using the precompute unit, which consists of $3^{G} = 27$ sets of adders and subtractors to cover all possible combinations of dynamic activations. The packed-up indexes are then used to address and select the corresponding values from the TL table. Finally, the selected values are accumulated to generate a single vector \( {\bf O} \in \mathbb{N}^{1 \times K} \) in the output matrix.

\subsubsection{Multi-TL-table dataflow design}


In our design, we carefully optimize the dataflow for TL-based matrix multiplication, as illustrated in Fig. \ref{fig:lutmatmul}. Assume we have a total of \( T \) tables. To better vectorize the TL-based matrix multiplication, the consecutive \( T \) indices can be grouped into an index vector, enabling simultaneous access to different look-up tables. The weight index vector matrix can be rewritten as \( W_{\text{idx}} \in \{5B, T\}^{\frac{N}{T\times G} \times K} \). \( W_{\text{idx}}\) are loaded onto the on-chip URAM.

Regarding the scheduling, the innermost loop first performs vector operations to establish the \( T \) look-up tables simultaneously, based on the first \( T \times G \) entries of the \( A \) matrix. Then, leveraging the multiple reading traits of the URAM \cite{URAM_Bind}, \( Q \) \( W \) index vectors are processed in parallel for TL table addressing, returning \( Q \times T \) outcomes. The corresponding TL table return values are then accumulated into an output buffer of size \( K \). The \( K \) index vectors on each row of \( W \) are traversed in steps of \( Q \). The TL table addressing and accumulation process can be fully pipelined with an interval of one cycle, as there are no inter-iteration dependencies. After the first \( T \times G \) entries of the \( A \) matrix are processed, the \( M \) values of the row of \( A \) are traversed in steps of \( T \times G \) in the intermediate loop. Finally, the outermost loop traverses the different row vectors in \( A \), corresponding to the tokens in the prefill stage of the LLM.

As for comparison, the LUT consumption of different matmul unit design methods is also presented in Table \ref{tab:lut_comparison}. The configuration for the TL-based matmul is set as \( G = 3 \), \( T = 32 \), and \( Q = 16 \). All levels of parallelism for the modules are set to be the same. Our design consumes 52094 LUTs, while the naive implementation, which selects whether to add or subtract, requires 7905 more LUTs. Another approach involves storing half of the possible combinations (13 out of 27) instead of all combinations and using the index to determine whether the value should be negative. This approach results in a smaller distributed RAM size, aiming to save LUTs. However, after synthesizing, it consumed 9209 more LUTs.

\begin{table}[h]
\centering
\caption{LUT Consumption Comparison of Different Matmul Unit Design Methods}
\resizebox{\linewidth}{!}{%
\begin{tabular}{|c|c|c|}
\hline
\textbf{Approach} & \textbf{LUT Consumption} & \textbf{Difference} \\

\hline
TL-based(Our Design) & 52,094 & -- \\
\hline
Naive Implementation & 59,999 & +7,905 \\
\hline
Partial Storage  & 61,303 & +9,209 \\
\hline

\end{tabular}%
}
\label{tab:lut_comparison}
\end{table}


\subsection{Specialized Reverse Reorder for Prefill Attention}
\begin{figure*}[t] 
\centering
\includegraphics[width=\linewidth]{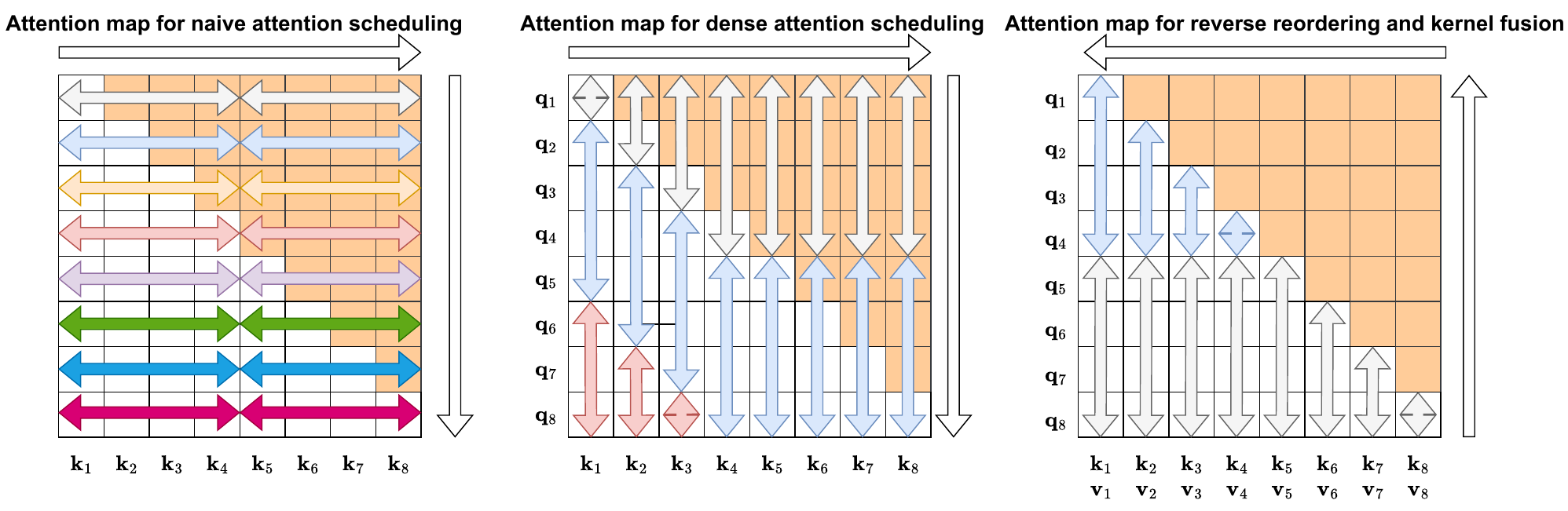}
\caption{The visualization of scheduling on the attention map (number of computation core $p = 4$, beige stands for attention mask).}
\label{fig:attention_map} 
\end{figure*}
\begin{figure}[t] 
\centering
\includegraphics[width=0.9\linewidth]{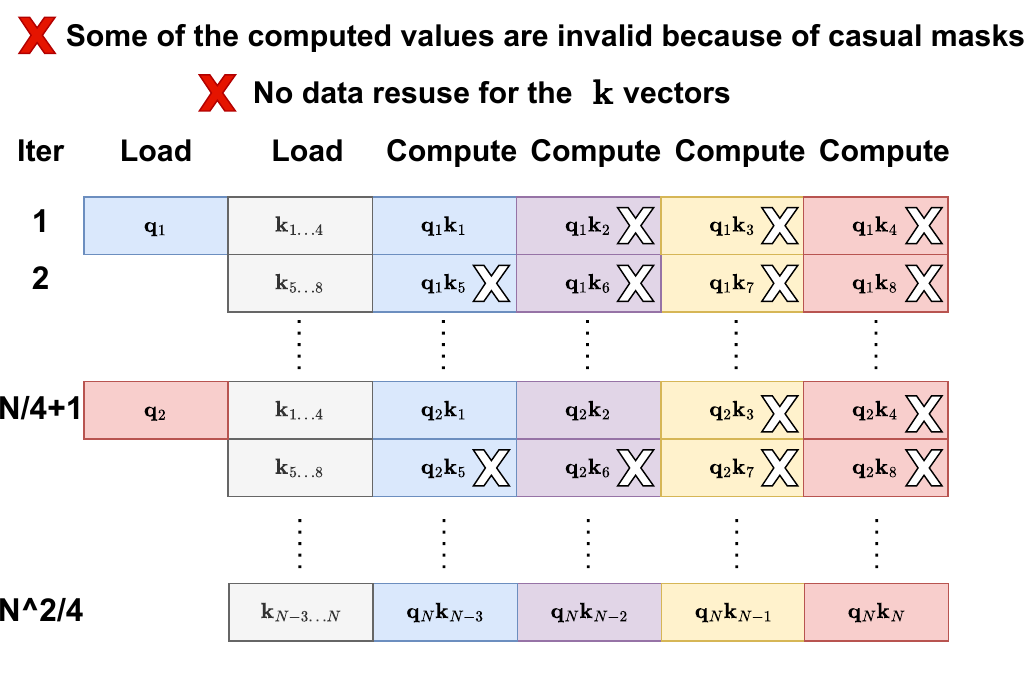}
\caption{Naive attention scheduling ($p = 4$).}
\label{fig:naive_schedule} 
\end{figure}

\subsubsection{Prefill Challenge on edge FPGA}
Prefill is one of the most challenging components for edge FPGAs. This part of LLM requires significant resources and bandwidth for multi-token computation, especially for attention computation, which involves softmax and matrix-to-matrix multi-head operations with a complexity of \( N^2 \). Given the limited memory bandwidth and finite computational units, the computation order of prefill attention must be carefully scheduled to meet these requirements. Otherwise, it may be constrained by the bandwidth limitations of the edge FPGA, as shown in the naive attention scheduling in Fig. \ref{fig:naive_schedule}. 

\begin{figure}[t] 
\centering
\includegraphics[width=0.9\linewidth]{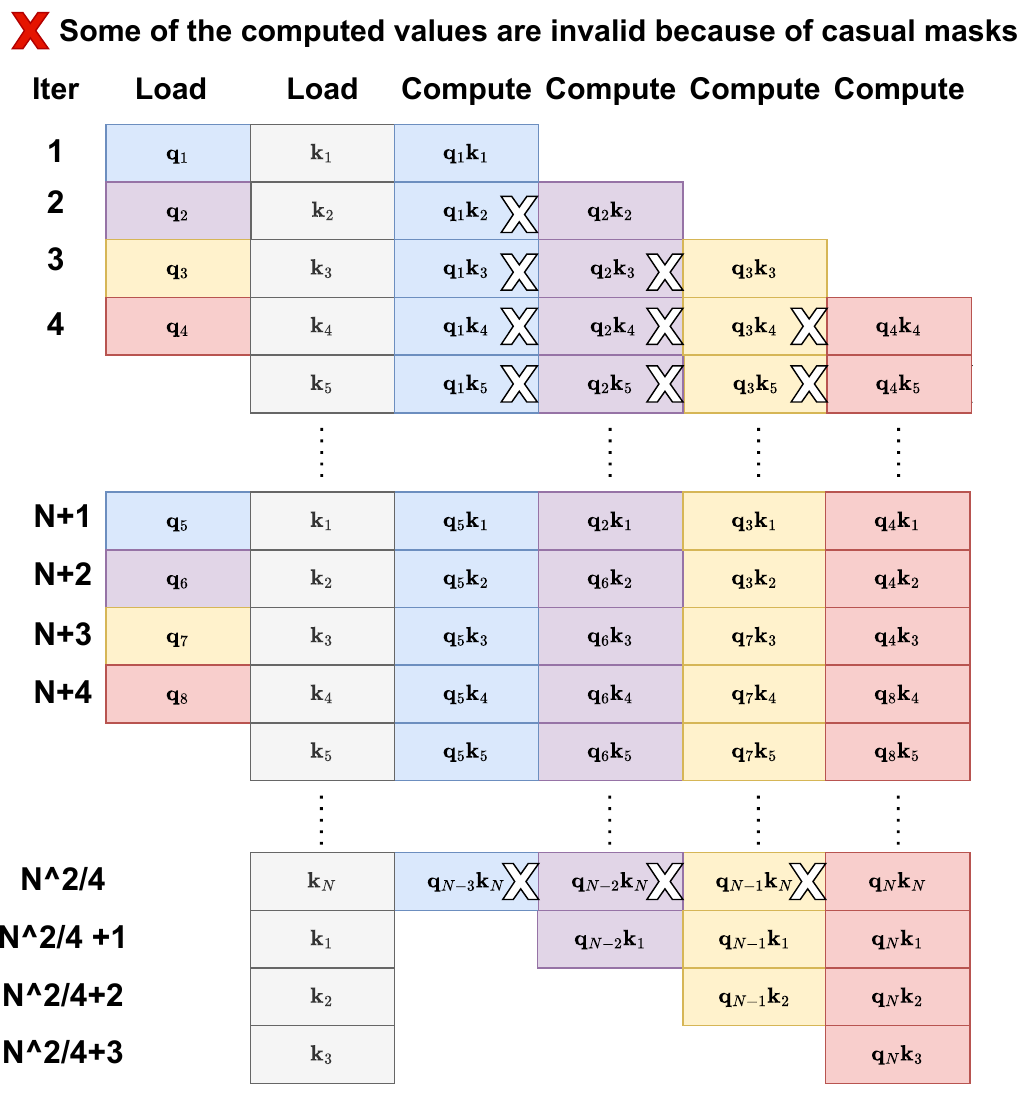}
\caption{Dense attention scheduling ($p = 4$).}
\label{fig:dense_schedule} 
\end{figure}

In the context of edge FPGA vision transformers, \cite{edgemoe} proposed a state-of-the-art dense attention scheduling strategy, as shown in Fig. \ref{fig:dense_schedule} and Fig. \ref{fig:attention_map}, which takes into account the reuse of the \( \bf Q \) values across different tokens. However, unlike vision transformers, LLMs use a causal attention mask. As a result, the dense scheduling approach wastes computational resources on zero masks.

Furthermore, the fusion of operations such as \( {\bf Q} \otimes {\bf K} \), softmax, and \( \bf S \otimes {\bf V} \) can reduce the additional accesses to DRAM. The state-of-the-art kernel fusion implementation for resource-abundant GPUs is Flash Attention. However, GPU-optimized computation is not suitable for FPGAs, as GPUs have many more computational cores and much larger on-chip SRAM compared to the on-chip BRAM/URAM available on FPGAs. To address these challenges, we propose the reverse attention method, which utilizes fused attention and reverse reorder scheduling, specifically tailored for edge FPGAs.

\begin{table}[ht]
\caption{Comparison of different attention approaches.}
\centering
\resizebox{\linewidth}{!}{
\begin{tabular}{|c|c|c|c|c|}
\hline
\textbf{Approach} & \textbf{Data Block Load} & \textbf{Iteration Count} & \textbf{Bandwidth} \\
\hline
Reverse Scheduling (Our Design)  & $\frac{N^2}{2p} + \frac{N}{2}  $ & $\frac{N^2}{2p} + \frac{N}{2}$ & $\sim 1$ \\
\hline
naive scheduling & $N^2 + N$ & $\frac{N^2}{p}$ & $\sim p$ \\
\hline
dense scheduling \cite{edgemoe}& $\frac{N^2}{p} + N + p - 1$ & $\frac{N^2}{p} + p - 1$ & $\sim 1$ \\

\hline
\end{tabular}
}
\label{tab:compare}
\end{table}

\begin{figure}[t] 
\centering
\includegraphics[width=\linewidth]{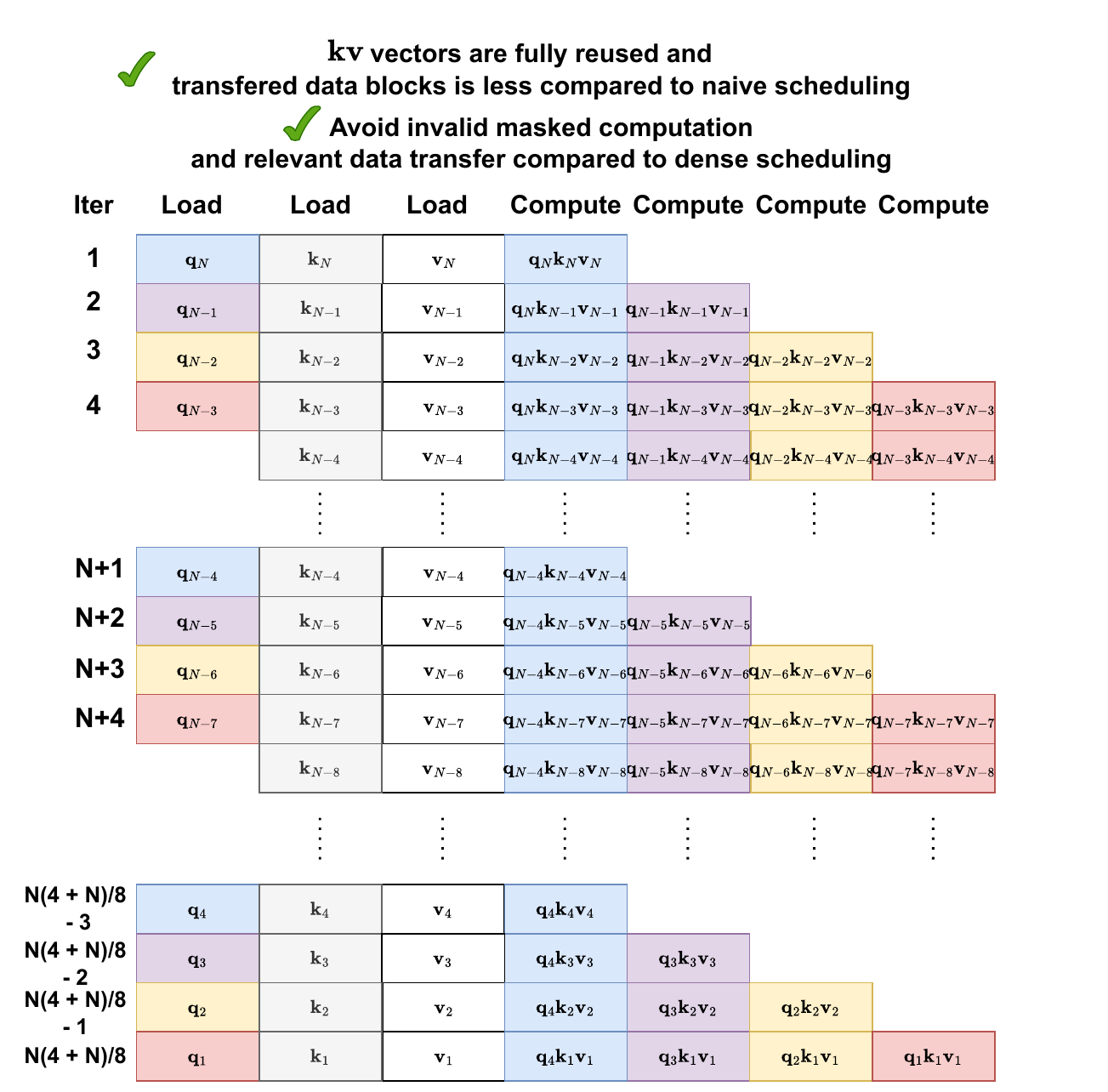}
\caption{Reverse attention scheduling ($p = 4$).}
\label{fig:reverse_schedule} 
\vspace{1em}
\end{figure}
\subsubsection{Reverse Attention}


The reverse attention scheduling is depicted in Fig. \ref{fig:reverse_schedule}. Assume that the current length of the prefill tokens is \( N \), with \( 1 < i \leq N \) and \( 1 < j \leq N \) representing the current token indices for \( \mathbf{q} \) and \( \mathbf{k}, \mathbf{v} \), respectively. There are a total of \( h \) heads.

The kernel fusion computation can be considered a special case of Flash Attention V2 \cite{dao2022flashattentionfastmemoryefficientexact} when the block size is equal to 1. The head-wise formula for the case with two consecutive blocks can be written as follows:
\begin{equation}
\left\{
\begin{aligned}
m^{(1)} &= s^{(1)} \\
\ell^{(1)} &= e^{s^{(1)}-m^{(1)}} \\
\mathbf{o}^{(1)} &= e^{s^{(1)}-m^{(1)}} \mathbf{v}^{(1)} \\
m^{(2)} &= \max \left(m^{(1)}, s^{(2)}\right) = m \\
\ell^{(2)} &= e^{m^{(1)}-m^{(2)}} \ell^{(1)} + e^{s^{(2)}-m^{(2)}} \\
&= e^{s^{(1)}-m} + e^{s^{(2)}-m} = \ell \\
\mathbf{p}^{(2)} &= e^{s^{(2)}-m^{(2)}} / \ell^{(2)} \\
\mathbf{o}^{(2)} &= \mathbf{o}^{(1)} / e^{m^{(1)}-m^{(2)}} + e^{s^{(2)}-m^{(2)}} \mathbf{v}^{(2)} \\
&= e^{s^{(1)}-m} \mathbf{v}^{(1)} + e^{s^{(2)}-m} \mathbf{v}^{(2)} \\
\mathbf{o}^{(2)} &= \mathbf{o}^{(2)} / \ell^{(2)} = \mathbf{o}
\end{aligned}
\right.
\tag{3}
\end{equation}
where \( m \) denotes the maximum value, \( s = \mathbf{q}_i \otimes \mathbf{k}_i \) represents the MAC result of the vector dot product, \( \ell \) is the denominator factor, and \( \mathbf{o} \) is the numerator vector. The upper index indicates the current step in the kernel fusion computation.

Regarding scheduling, instead of starting from the first token \( \mathbf{q}_1 \), our schedule begins from \( \mathbf{q}_{N-1} \). Specifically, the level of parallelism is set to \( p \) (as illustrated in the figure for \( p = 4 \)). The factor \( p \) also implies that the on-chip BRAM can store \( p \) tokens of \( \mathbf{q}_i \). In each iteration, one \( \mathbf{q}_i \) token is loaded onto the on-chip memory (with the first batch loading \( \mathbf{q}_N \) to \( \mathbf{q}_{N-3} \)). Simultaneously, the corresponding \( \mathbf{k}_j \) and \( \mathbf{v}_j \) tokens are loaded for computation. After all \( N \) \( \mathbf{k}_j \) and \( \mathbf{v}_j \) tokens have been loaded and the fused-kernel computation is completed, the next iteration will evict \( p \) \( \mathbf{k}_j \) and \( \mathbf{v}_j \) tokens, starting from \( \mathbf{k}_{N-3} \) and \( \mathbf{v}_{N-3} \), to avoid redundant computations arising from the causal attention mask. The iteration continues until all \( 1 < i \leq N \), \( 1 < j \leq N \) are traversed. In this approach, the only required input buffers are for \( p \) \( \mathbf{q}_i \) tokens, one \( \mathbf{k}_j \), and one \( \mathbf{v}_j \). Additionally, the intermediate buffers include: \( h \times p \) multi-head MAC intermediate results \( s \), \( h \times p \) multi-head previous max values \( m \), and \( h \times p \) intermediate denominators \( \ell \).

After introducing the reverse attention process, a comparison between naive attention, dense attention in Edge Moe \cite{edgemoe}, and our proposed reverse attention is provided in Table \ref{tab:compare}. The comparison indicates that the iteration count for reverse attention is the lowest, and the required bandwidth remains constant. Furthermore, the redundant outcome rate of the computation core can be directly assessed from the attention map in Fig. \ref{fig:attention_map}. The naive and dense scheduling approaches do not account for the causal mask, leading to many redundant computed values. 
\subsection{Hardware Specialization and Reuse for Decoding-Phase Attention and LM Head}

The previous section focuses on optimizing the Prefill phase in LLM inference. Since we consider efficient end-to-end LLM inference on single edge Device, both Prefill and Decoding phases require efficient implementation to maximize global performance.
The attention mechanism is a core component in both phases, but its computational characteristic differs. In the prefill phase, calculating attention involves operations on matrices representing the entire input sequence. In the decoding phase, it involves operations between the vector representation of the single new token and the cached matrices of keys and values. This difference presents an opportunity for hardware specialization.

\begin{figure}[ht]
    \centering

    \includegraphics[width=\linewidth]{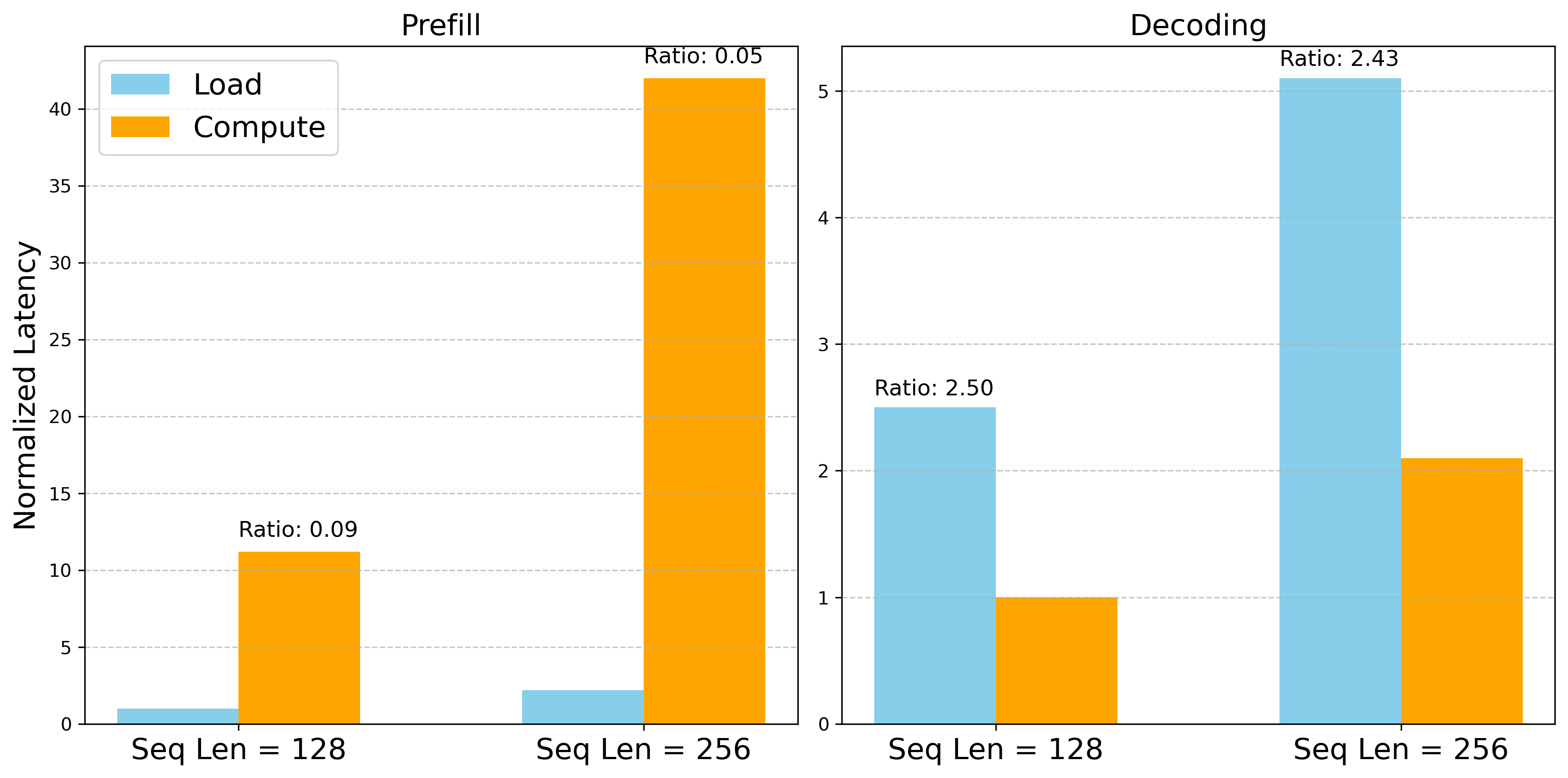}

    \caption{Characterization of Attention Module during Prefill/Decoding Phase }
    \label{fig:load_compute}
\end{figure}

In the decoding phase, a single new token is generated per step. Let the total sequence length (prompt + already generated tokens) be $M$. The query $\bf q$ is now a $1 \times N$ vector corresponding to the new token. The key ($K_{cache}$) and value ($V_{cache}$) matrices contain the cached representations of all M previous tokens and are of dimensions M×d. The core attention computation involves:

1.	Calculating attention scores: ${\bf q} \times K_{cache}^T$ (a $1 \times N$ vector multiplied by a $N\times M$ matrix, resulting in a ${1\times M}$ score vector).
2.	Applying softmax to the scores.
Multiplying the resulting $1\times M$ vector by $V_{cache}$ (an $M\times N$ matrix) to get the $1\times N$ output vector.
The computation involves primarily matrix-vector and vector-vector operations. The computational load per step ($O(Nd)$) is significantly lower than the total prefill computation. However, this phase requires fetching the large $K_{cache}$ and $V_{cache}$ matrices from memory (e.g., off-chip DRAM) in every step. Consequently, the decoding phase is often memory-bandwidth bound, especially as the sequence length $M$ grows. 
Since computation is less intensive and latency is dominated by memory access (fetching the KV cache), massive parallelism is inefficient and wastes resources. A more sequential or lower-parallelism computation unit is sufficient. This approach significantly reduces the required on-chip hardware resources (e.g., number of PEs, buffer sizes) compared to the prefill unit. 
The profiling result in Figure~\ref{fig:load_compute} clearly shows that the attention in the decoding phase is memory-bounded, while the prefill phase is compute-bounded. This demonstrates the necessity of lightweight decoding implementation to save on-chip resources for the Prefill phase.

\subsubsection{Reuse of Decoding Attention for LM Head}
Following the processing 
thr-ough N transformer blocks in typical LLM architectures like LLaMA, the final step before token generation involves the LM Head. This component performs a crucial linear projection, mapping the final hidden state output from the last transformer block to a vector of logits representing the probability distribution over the entire vocabulary.
For inference, particularly during the auto-regressive decoding phase, where one token is generated at a time, the input to the LM Head is the final hidden state vector corresponding to the token being predicted. This vector has dimensions $[1,N]$, where N represents the hidden state dimension (e.g., $N=1536$ for some models). The LM Head uses a large weight matrix of dimensions $[N,V]$, where $V$ is the vocabulary size (e.g., $V=32000$), to compute the output logit vector of dimensions $[1,V]$. The core computation is thus a matrix-vector multiplication: $[1,N]\times[N,V]\rightarrow [1,V]$.
 This computation (O(HV)) has similar characteristics to decoding attention: it's a matrix-vector operation with limited data reuse opportunities, heavily reliant on fetching a large matrix (the $K_{cache}$ or the LM Head weights), making it memory-bound (especially as V $\gg$ H).
 
Given this similarity, we reuse the decoding-phase attention hardware to execute the LM Head computation. This eliminates the need for a dedicated LM Head unit, yielding substantial area and power savings. The performance impact is negligible because the LM Head executes only once per generated token, whereas attention occurs N times (once per layer).

Reusing the Decoding-phase Attention hardware for both attention and the LM Head precludes a fully fused attention pipeline within it. We therefore adopt a decoupled execution model for attention sub-steps during decoding:
(1) Attention Score Computation: ${\bf s}={\bf q}\times K_{cache}^T$ .
(2) Softmax: ${\bf p}=softmax({\bf s})$.
(3) Value Aggregation: Compute the final attention output ${\bf o}={\bf p}\times V_{cache}$.
This is efficient because the intermediate score/probability vector (1×M) is small enough to be buffered on-chip BRAM with minimal latency penalty.
On the contrary, for the Prefill phase, the intermediate N×N attention score matrix would be far too large for practical on-chip buffering. Thus it requires a fully fused attention pipeline that integrates attention score calculation, softmax, and value aggregation in a single, uninterrupted hardware pass to minimize off-chip data movement.

\subsection{Implementation and Optimization of Special Function Units}

Besides the core modules discussed in previous sections, LLM inference also relies on several essential special functions, including Quantization / Dequantization, RMSNorm, and Activation Function. 
These operations are computationally less intensive, therefore, our strategy focuses on lightweight hardware implementations and operator fusion to minimize overhead. For efficient data handling, vector data for these modules is processed in 256-bit packets, aligning with AXI bus width.

\textbf{Quantization \& Dequantization}: 
The activations need to be quantized before the ternary Linear modules. We employ Absmax Quantization, which involves two passes: (1) finding the maximum absolute value to compute the scale factor, and (2) applying this scale element-wise. Figure~\ref{Fig:Bitnet} shows that quantization follows the RMSNorm. We fuse these operations to reduce data movement. The dequantization is fused into the Linear output pipeline.

\textbf{RMSNorm}:
RMSNorm involves two passes. The first calculates the Root Mean Square of the input x: $RMS(x) =\sqrt{\frac{1}{n}\sum_{i =1}^{N}x_{i}^{2}}$. The second pass normalizes the input by dividing by the RMS value and multiplying by a learned scaling parameter $\gamma$. Recognizing that both RMSNorm and Absmax Quantization involve a two-pass traverse, we fuse these four logical steps into two optimized hardware passes. This significantly minimizes the data movement.

\textbf{Activation function}:
The element-wise SiLU activation ($x\cdot\frac{1}{1+e^{-x}}$) is required after the Gate projection in Feed-Forward Network (FFN) block. The SiLU is pipelined and fused directly into the preceding Linear module, effectively hiding its latency.

\begin{table*}[t]
\centering
\caption{Comparison of FPGA-based LLM Accelerators}
\label{tab:fpga-comparison}
\resizebox{\textwidth}{!}{%
\begin{tabular}{@{}lccccccccccc@{}}
\toprule
\textbf{Work} & \textbf{Device} & \textbf{LUT} & \textbf{FF} & \textbf{BRAM} & \textbf{DSP} & \textbf{MHz} & \textbf{Power (W)} & \textbf{BW (GB/s)} & \textbf{Model} & \textbf{Throughput (tokens/s)} & \textbf{Accelerate Prefill?} \\ 
\midrule
SECDA~\cite{haris2024designing}        & PYNQ     & --   & --    & --   & --   & --   & --   & --    & TinyLLaMA W4 & 0.58 & \ding{55} \\
LlamaF~\cite{llamaf}      & ZCU102   & 164K & 171K  & 223  & 528  & 205  & 5.08 & 21.3  & TinyLLaMA W8    & 1.50 & \ding{55}  \\
Li et al.~\cite{li2025pushing}& KV260    & 78K  & 105K  & 36.5 & 291  & 300  & 6.57 & 19.2  & LLaMA2-7B W4     & 4.90 & \ding{55} \\ \hline
\textbf{TeLLMe(Ours)}& KV260    & 108K$^*$  & 155K$^*$  & 206$^*$ & $356^*$  & $250^*$  & $6.72^*$ & 19.2  & Bitnet W1.58     & 9.51 & \ding{51} \\
\hline
\end{tabular}
}
\vspace{2mm}
\caption*{\footnotesize{* Not directly comparable since our TeLLMe have additional logic to accelerate prefill stage}}
\end{table*}

\section{Experimental Results and Analysis} \label{expr}
\subsection{Experiment Setup}
We implement the TeLLMe accelerator using high-level synthesis C/C++ in Vitis HLS  and Vivado 2023.1. We evaluate our design on Kria KV260 (Zynq UltraScale+ XCK26 MPSoC). To ensure timing closure, we use 250 MHz for the final bitstream generation.
\subsection{LLM Inference Performance and Resource Breakdown}
\begin{figure}[ht]
    \centering

    \includegraphics[width=\linewidth]{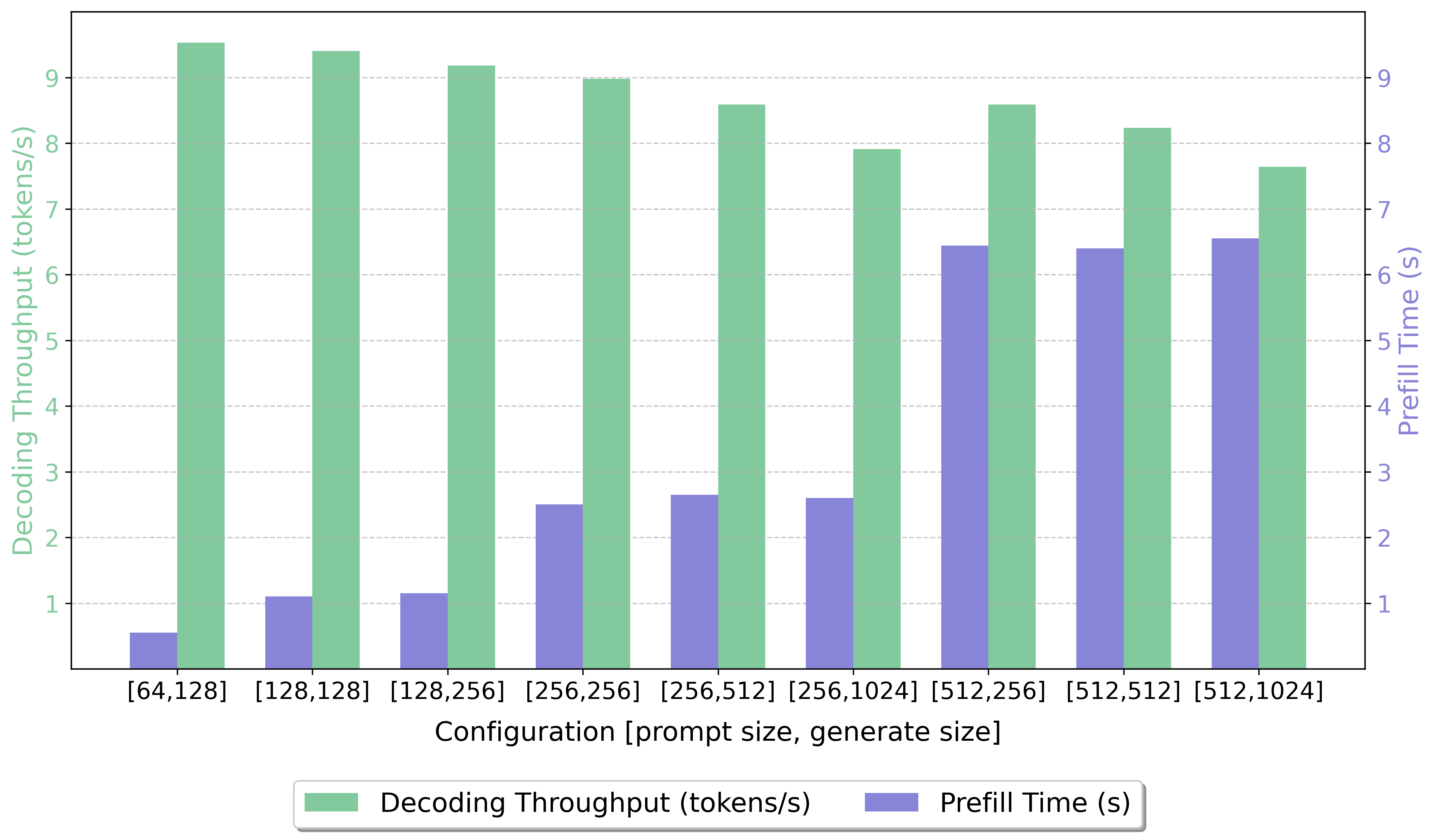}

    \caption{TeLLMe LLM Inference Performance}
    \label{fig:perf}
    \vspace{4mm}
\end{figure}

Figure~\ref{fig:perf} shows the key metrics in LLM inference, including Decoding Throughput (token generation speed) and Prefill Time (time-to-first-token). We evaluate TeLLMe under different configurations [prompt size, generate size], note that the total tokens = prompt size + generate size. TeLLMe achieves $>$ 9 tokens/s in 512 context lengths and \textasciitilde8 tokens/s in 1024 context lengths. For prompt size $<$ 128, TeLLMe achieves \textasciitilde1s Prefill size. TeLLMe demonstrates practical viability and deployment potential in real-world
applications.

The resource breakdown is shown in Table \ref{tab:utilization}. Regarding BRAM usage, most of it is consumed by the top-level AXI buffer. DSP resources are mainly utilized by the Attention modules due to their INT8 precision. LUTs are primarily consumed by the TL-based matmul unit for the TL tables. URAM is used by the matmul weight buffer to support ping-pong operations.

\begin{table}[h]
\centering
\caption{Resource Consumption Breakdown}
\label{tab:utilization}
\resizebox{\columnwidth}{!}{%
\begin{tabular}{cccccc}
    \toprule
Module                     & BRAM & DSP & FF    & LUT   & URAM \\ \hline
Control \& Data Transfer   & 120       & 0   & 24973 & 5897  &     \\
Attention (Prefill Phase)  & 46        & 122 & 25629 & 33069 &      \\
Attention (Decoding Phase) & 24        & 134 & 17465 & 7028  &     \\
TL-based Matmul Unit   & 0         & 0   & 35765 & 52094 & 48   \\
RMSNorm                    & 16        & 28  & 6202  & 5933  &      \\
Misc (Add, Mul, RoPE, etc) & 0         & 72  & 45804 & 4973  &     \\ \hline
\multirow{2}{*}{\textbf{Total}} & 206  & 356    & 155838  & 108994  & 48\\
\multirow{2}{*}{}               & (71\%)& (28\%)& (66\%)      & (93\%)   & (75\%)  \\     \bottomrule
\end{tabular}%
}
\end{table}

\subsection{Comparison with Existing Edge FPGA Work}

Table~\ref{tab:fpga-comparison} presents a comparison between TeLLMe and prior FPGA-based LLM accelerators. Despite differences in model scale and quantization schemes, TeLLMe achieves a peak decoding throughput of 9.51 tokens per second—representing up to a \textbf{$16.4\times$ improvement} over previous work—while supporting both prefill and decoding stages on a single edge FPGA device. As highlighted in the table, none of the existing edge FPGA-based solutions implement on-device prefill, often citing its computational intensity as unsuitable for resource-constrained hardware. While we acknowledge the challenges associated with prefill on FPGAs, we argue that full on-device support is essential for a complete and self-contained edge deployment. Relying on external hosts to perform prefill introduces additional concerns regarding system complexity, data privacy, and scalability. Our detailed prefill performance is presented in Figure~\ref{fig:perf} and Table~\ref{tab:prefill_results}.
\subsection{Comparison with Mobile CPU}
\renewcommand{\arraystretch}{1.15}
\begin{table}[t]
\centering
\caption{Performance Comparison with Mobile CPU}
\resizebox{\columnwidth}{!}{%
\begin{tabular}{l|l|c|c|c|c}
\hline
\textbf{Device} & \textbf{Category} & \begin{tabular}[c]{@{}c@{}}Decode\\ (tokens/s)\end{tabular} & \begin{tabular}[c]{@{}c@{}}Time-to-\\ first-token (s)\end{tabular} & \begin{tabular}[c]{@{}c@{}}Prefill\\ (tokens/s)\end{tabular} & \begin{tabular}[c]{@{}c@{}}Model Size\\ (MB)\end{tabular}  \\
\hline
\multirow{6}{*}{\shortstack{Snapdragon\\8 Gen 3}} 
& 1B BF16 (baseline)   & 19.2  & 1.0  & 60.3  & 2358 \\
& 1B SpinQuant         & 50.2  & 0.3  & 260.5 & 1083 \\
& 1B QLoRA             & 45.8  & 0.3  & 252.0 & 1127 \\
& 3B BF16 (baseline)   & 7.6   & 3.0  & 21.2  & 6129 \\
& 3B SpinQuant         & 19.7  & 0.7  & 89.7  & 2435 \\
& 3B QLoRA             & 18.5  & 0.7  & 88.8  & 2529 \\
\hline
\textbf{KV260 FPGA} & \textbf{0.7B TeLLMe} & \textbf{9.51} & \textbf{0.55} & \textbf{116.4} & \textbf{257} \\
\hline
\end{tabular}
}
\vspace{2mm}
\caption*{\footnotesize{* Time-to-first-token (prefill delay) is measured with a prompt length = 64 }}
\label{tab:prefill_results}
\end{table}
Despite the significant technological disparity between KV260 and modern mobile SoCs, our TeLLMe design demonstrates highly competitive performance in key inference metrics. As shown in Table~\ref{tab:prefill_results}, TeLLMe achieves a prefill latency of 0.55 seconds—comparable to the 0.3–0.7 seconds observed on the Qualcomm Snapdragon 8 Gen 3, a device fabricated in an advanced 4nm process with integrated LPDDR5x memory and substantially higher bandwidth. In contrast, the KV260 is based on a 16nm process and relies on DDR4 memory with much lower bandwidth. This makes our ability to match prefill performance particularly notable, as prefill is typically compute-bound and less amenable to acceleration on bandwidth-constrained FPGAs. This result highlights the effectiveness of our architectural optimizations—including TL-based tenary matmul, a bandwidth-efficient attention module with fused
operation and a Reversed Attention reordering scheme to accelerate prefill.

While TeLLMe’s decoding throughput (9.51 tokens/s) lags behind that of mobile SoCs, this gap is largely attributable to the KV260's limited external memory bandwidth, which disproportionately affects the memory-bound decode phase. We emphasize that this limitation is architectural rather than algorithmic; our design scales favorably to higher-bandwidth platforms such as HBM-enabled FPGAs or custom ASICs. Taken together, these results validate TeLLMe as the first binary LLM accelerator on edge FPGA to support full inference—including both prefill and decoding—with energy efficiency and architectural flexibility that position it well for future edge AI deployments.
\section{Conclusion}
We introduced \textbf{TeLLMe}, the first end-to-end FPGA accelerator optimized for ternary LLM inference across both prefill and decoding stages. By co-optimizing compute, memory, and scheduling, TeLLMe employs a table-lookup-based matmul engine that reuses grouped activations and online precomputations across projection and feedforward layers for efficient ternary matrix operations. A fused attention module with reversed attention and Flash Attention-style kernel fusion reduces bandwidth demands, eliminates redundant masked operations, and supports parallelism. Running under 7 W, TeLLMe achieves up to 9.51 tokens/s and supports 1024-token contexts, outperforming mobile SoCs at significantly lower power. It delivers prefill latencies of 0.55–1.15 s for prompts of 64–128 tokens. To our knowledge, TeLLMe is the first real-hardware FPGA accelerator to fully support ternary LLMs end-to-end, establishing a new benchmark for efficient, low-latency edge inference.
\bibliography{refernces.bib}{}

\begin{thebibliography}{10}
\providecommand{\url}[1]{#1}
\csname url@samestyle\endcsname
\providecommand{\newblock}{\relax}
\providecommand{\bibinfo}[2]{#2}
\providecommand{\BIBentrySTDinterwordspacing}{\spaceskip=0pt\relax}
\providecommand{\BIBentryALTinterwordstretchfactor}{4}
\providecommand{\BIBentryALTinterwordspacing}{\spaceskip=\fontdimen2\font plus
\BIBentryALTinterwordstretchfactor\fontdimen3\font minus \fontdimen4\font\relax}
\providecommand{\BIBforeignlanguage}[2]{{%
\expandafter\ifx\csname l@#1\endcsname\relax
\typeout{** WARNING: IEEEtran.bst: No hyphenation pattern has been}%
\typeout{** loaded for the language `#1'. Using the pattern for}%
\typeout{** the default language instead.}%
\else
\language=\csname l@#1\endcsname
\fi
#2}}
\providecommand{\BIBdecl}{\relax}
\BIBdecl

\bibitem{brown2020language}
T.~Brown, B.~Mann, N.~Ryder, M.~Subbiah, J.~D. Kaplan, P.~Dhariwal, A.~Neelakantan, P.~Shyam, G.~Sastry, A.~Askell \emph{et~al.}, ``Language models are few-shot learners,'' \emph{Advances in neural information processing systems}, vol.~33, pp. 1877--1901, 2020.

\bibitem{touvron2023llama}
H.~Touvron, T.~Lavril, G.~Izacard, X.~Martinet, M.-A. Lachaux, T.~Lacroix, B.~Rozi{\`e}re, N.~Goyal, E.~Hambro, F.~Azhar \emph{et~al.}, ``Llama: Open and efficient foundation language models,'' \emph{arXiv preprint arXiv:2302.13971}, 2023.

\bibitem{guo2025deepseek}
D.~Guo, D.~Yang, H.~Zhang, J.~Song, R.~Zhang, R.~Xu, Q.~Zhu, S.~Ma, P.~Wang, X.~Bi \emph{et~al.}, ``Deepseek-r1: Incentivizing reasoning capability in llms via reinforcement learning,'' \emph{arXiv preprint arXiv:2501.12948}, 2025.

\bibitem{qiao2022two}
Y.~Qiao, M.~Alnemari, and N.~Bagherzadeh, ``A two-stage efficient 3-d cnn framework for eeg based emotion recognition,'' in \emph{2022 IEEE International Conference on Industrial Technology (ICIT)}.\hskip 1em plus 0.5em minus 0.4em\relax IEEE, 2022, pp. 1--8.

\bibitem{10025006}
A.~Ding, Y.~Qiao, and N.~Bagherzadeh, ``Bnn an ideal architecture for acceleration with resistive in memory computation,'' \emph{IEEE Transactions on Emerging Topics in Computing}, vol.~11, no.~2, pp. 281--291, 2023.

\bibitem{bitnet}
H.~Wang, S.~Ma, L.~Dong, and et~al., ``Bitnet: Scaling 1-bit transformers for large language models,'' \emph{arXiv preprint arXiv:2310.11453}, 2023.

\bibitem{bitnet158}
S.~Ma, H.~Wang, L.~Ma, and et~al., ``The era of 1-bit llms: All large language models are in 1.58 bits,'' \emph{arXiv preprint arXiv:2402.17764}, 2024.

\bibitem{deepseek}
D.~Guo, D.~Yang, H.~Zhang, and et~al., ``Deepseek-r1: Incentivizing reasoning capability in llms via reinforcement learning,'' \emph{arXiv preprint arXiv:2501.12948}, 2025.

\bibitem{li2025pushing}
J.~Li, T.~Li, G.~Shen, D.~Zhao, Q.~Zhang, and Y.~Zeng, ``Pushing up to the limit of memory bandwidth and capacity utilization for efficient llm decoding on embedded fpga,'' \emph{arXiv preprint arXiv:2502.10659}, 2025.

\bibitem{chen2024understanding}
H.~Chen, J.~Zhang, Y.~Du, S.~Xiang, Z.~Yue, N.~Zhang, Y.~Cai, and Z.~Zhang, ``Understanding the potential of fpga-based spatial acceleration for large language model inference,'' \emph{ACM Transactions on Reconfigurable Technology and Systems}, vol.~18, no.~1, pp. 1--29, 2024.

\bibitem{fbillm}
L.~Ma, M.~Sun, and Z.~Shen, ``Fbi-llm: Scaling up fully binarized llms from scratch via autoregressive distillation,'' \emph{arXiv preprint arXiv:2407.07093}, 2024.

\bibitem{onebit}
Y.~Xu, X.~Han, Z.~Yang, and et~al., ``Onebit: Towards extremely low-bit large language models,'' in \emph{Advances in Neural Information Processing Systems (NeurIPS)}, 2024.

\bibitem{bitdistiller}
D.~Du, Y.~Zhang, S.~Cao, and et~al., ``Bitdistiller: Unleashing the potential of sub-4-bit llms via self-distillation,'' in \emph{Proceedings of the 62nd Annual Meeting of the Association for Computational Linguistics (ACL)}, 2024.

\bibitem{quip}
J.~Chee, Y.~Cai, V.~Kuleshov, and C.~De~Sa, ``Quip: 2-bit quantization of large language models with guarantees,'' \emph{arXiv preprint arXiv:2307.13304}, 2023.

\bibitem{tmac}
J.~Wei, S.~Cao, T.~Cao, and et~al., ``T-mac: Cpu renaissance via table lookup for low-bit llm deployment on edge,'' \emph{arXiv preprint arXiv:2407.00088}, 2024.

\bibitem{llamaf}
H.~Xu, Y.~Li, and S.~Ji, ``Llamaf: An efficient llama2 architecture accelerator on embedded fpgas,'' \emph{arXiv preprint arXiv:2409.11424}, 2024.

\bibitem{edgemoe}
R.~Sarkar, H.~Liang, Z.~Fan, Z.~Wang, and C.~Hao, ``Edge-moe: Memory-efficient multi-task vision transformer architecture with task-level sparsity via mixture-of-experts,'' in \emph{IEEE/ACM International Conference on Computer-Aided Design (ICCAD)}, 2023, pp. 1--9.

\bibitem{haris2024designing}
\BIBentryALTinterwordspacing
J.~Haris, R.~Saha, W.~Hu, and J.~Cano, ``Designing efficient llm accelerators for edge devices,'' \emph{arXiv preprint arXiv:2408.00462}, 2024, accessed: 2025-04-20. [Online]. Available: \url{https://arxiv.org/abs/2408.00462}
\BIBentrySTDinterwordspacing

\bibitem{TLMAC}
D.~Gerlinghoff, B.~Choong, R.~Goh, W.-F. Wong, and T.~Luo, ``Table-lookup mac: Scalable processing of quantised neural networks in fpga soft logic,'' 04 2024, pp. 235--245.

\bibitem{URAM_Bind}
\BIBentryALTinterwordspacing
AMD, ``Uram storage bind,'' 2024. [Online]. Available: \url{https://docs.amd.com/r/2024.2-English/ug1399-vitis-hls/pragma-HLS-bind\_storage}
\BIBentrySTDinterwordspacing

\bibitem{dao2022flashattentionfastmemoryefficientexact}
\BIBentryALTinterwordspacing
T.~Dao, D.~Y. Fu, S.~Ermon, A.~Rudra, and C.~Ré, ``Flashattention: Fast and memory-efficient exact attention with io-awareness,'' 2022. [Online]. Available: \url{https://arxiv.org/abs/2205.14135}
\BIBentrySTDinterwordspacing

\end{thebibliography}

\bibliographystyle{IEEEtran}

\end{document}